\documentclass[11pt,twoside]{article}
\usepackage{asp2010}
\resetcounters

\begin{document}

\title{Characterizing Ultraviolet Excesses in the Outskirts of a Local Early-Type Sample}
\author{Jennifer Donovan Meyer$^{1,2}$, J.~H. van Gorkom$^2$, and David Schiminovich$^2$
\affil{$^1$Department of Physics \& Astronomy, Stony Brook University, Stony Brook, NY}
\affil{$^2$Department of Astronomy, Columbia University, New York, NY}}

\begin{abstract}
We present an analysis of ultraviolet (UV) emission in the outer regions of a local, volume-limited sample of 56 early-type galaxies, where H$\alpha$ emission from massive star formation is typically absent. We find excess faint NUV emission in the environments of our early-type galaxies compared to blank sky measured in the same tiles, indicating that the excesses are not due to background contamination. We do not observe corresponding faint FUV excesses. Faint NUV excesses increase with galaxy luminosity and are not correlated with the presence or absence of HI in the environments of these galaxies. The faint NUV excesses in the outskirts of early-type galaxies can be interpreted as being due to star formation at or above a few $\times$ 10$^{-5}$ M$_{\odot}$ yr$^{-1}$ kpc$^{-2}$; star formation at this rate can create a few percent of the mass of an early-type galaxy in a Gyr. Faint early types (with M$_{B}$ $>$ -21.3) have on average four times as many bright UV sources within 30 kpc compared to bright early types (with M$_{B}$ $<$ -21.3). The peak of the source distribution detected around faint early types is less luminous and slightly bluer than the peak of the sources detected around bright early types, indicating that early types with M$_{B}$ $>$ -21.3 are more actively building up their mass with young stars. The spatial distribution of bright sources around all early types increases approximately linearly out to 20 kpc and subsequently flattens. 
\end{abstract}

\section{Introduction}

Determining whether recent star formation exists within early-type galaxies or their environments is an important test for theories of the evolution of these systems. By constraining the extent to which star formation can be responsible for the buildup of mass in red sequence galaxies at late times -- a factor of $\sim$2 increase since z=1 is observed (i.e., \citealt{Bell04}) -- limits can be placed on how important other evolutionary paths (i.e., gasless ``dry" merging) may be. 

Moderate levels of recent star formation traced by blue UV colors are found at the centers of some early types \citep{Jeong09}, accompanied in some cases by dense disks of molecular gas \citep{Crocker10}, but the generally hot centers of early types where cold gas has settled are not the only places to find recently formed stars in and around early-type galaxies. Extended, structured ultraviolet emission has been detected in the outer regions of individual early-type galaxies \citep{Donovan09, Jeong07, Thilker10}, but the statistical significance of this phenomenon within the early-type galaxy population has not yet been established. 

However, the statistical significance of neutral hydrogen in and around early-type galaxies in the local universe $\it{is}$ currently being established. Atlas 3D will present the first complete, unbiased survey of local early types observed in HI; currently this project has detected roughly 50\% of the local field early type population down to a few x 10$^{7}$ M$_{\odot}$ in a sample of 260 galaxies \citep{Serra2010}, the same percentage found by \citet{vanGorkom97} in a smaller, more heterogeneous sample. Such a high detection rate of gas in and around these galaxies argues for corresponding searches for star formation associated with these systems, and accordingly, a study of the IMF in such extreme, low $\Sigma_{SFR}$ environments. It would be interesting to determine whether star formation in the outer regions of early types shows the same lack of H$\alpha$ emission, and hence massive star formation, as has been found in the outer regions of many spirals.

The sample of systems presented in this work, partnered with the low background in the GALEX images, for the first time reveals the prevalence of excess UV emission distributed in the outskirts of the optical bodies of early-type galaxies. We search for faint UV excesses at large galactic radius (out to 30 kpc from each galaxy) and study the distribution of bright UV sources extracted from the same regions. We also investigate whether the presence or absence of HI is correlated with these excesses. 

\section{Observations}

We present the ultraviolet properties of a volume-limited sample of E and S0 galaxies selected by \citet{Schweizer92}; of the 69 galaxies in their sample, 56 have been observed by GALEX. The sample was selected for being bright (m$_{B}$ $\leq$ 13.5) and north of $\delta$=-15$^{\circ}$ (making the galaxies visible from the VLA). The selection criteria also required the systems to be close (v $<$ 4000 km s$^{-1}$), making the resolution sufficient to discern the distribution of star formation in these galaxies. Assuming H$_{o}$=70 km s$^{-1}$ Mpc$^{-1}$, the 5'' resolution of the GALEX detectors corresponds to a spatial size of $\sim$1.4 kpc at the volume limit of 57.1~Mpc. 

Each observed galaxy was imaged for 1500 seconds in both the NUV and FUV bands, though a few galaxies have significantly more or slightly less exposure time. HI images have been published in the literature for a subsample of the galaxies.

\section{Results}

\noindent $\it{Faint~UV~Excesses:}$ In Figure~1, we show the average counts per pixel (expressed in magnitudes per square arcsecond) within 30 kpc of each galaxy as a function of the average counts per pixel in the background. Sources detected at or above 5$\sigma$ -- including the central galaxy -- are removed before computing pixel averages. A dot-dashed line with slope equal to 1.0 is overlaid on both plots; any system where the average counts within 30 kpc of the galaxy equal the average counts in the background would appear as a point on this line. In the NUV plot, we also show a linear best fit to the data. There is a NUV excess inconsistent with being due to background sources in the environments of local early type galaxies which is comprised of diffuse emission and/or NUV sources detected at less than 5$\sigma$ by GALEX. Such an excess does not exist in the FUV. 

\begin{figure}
\plottwo{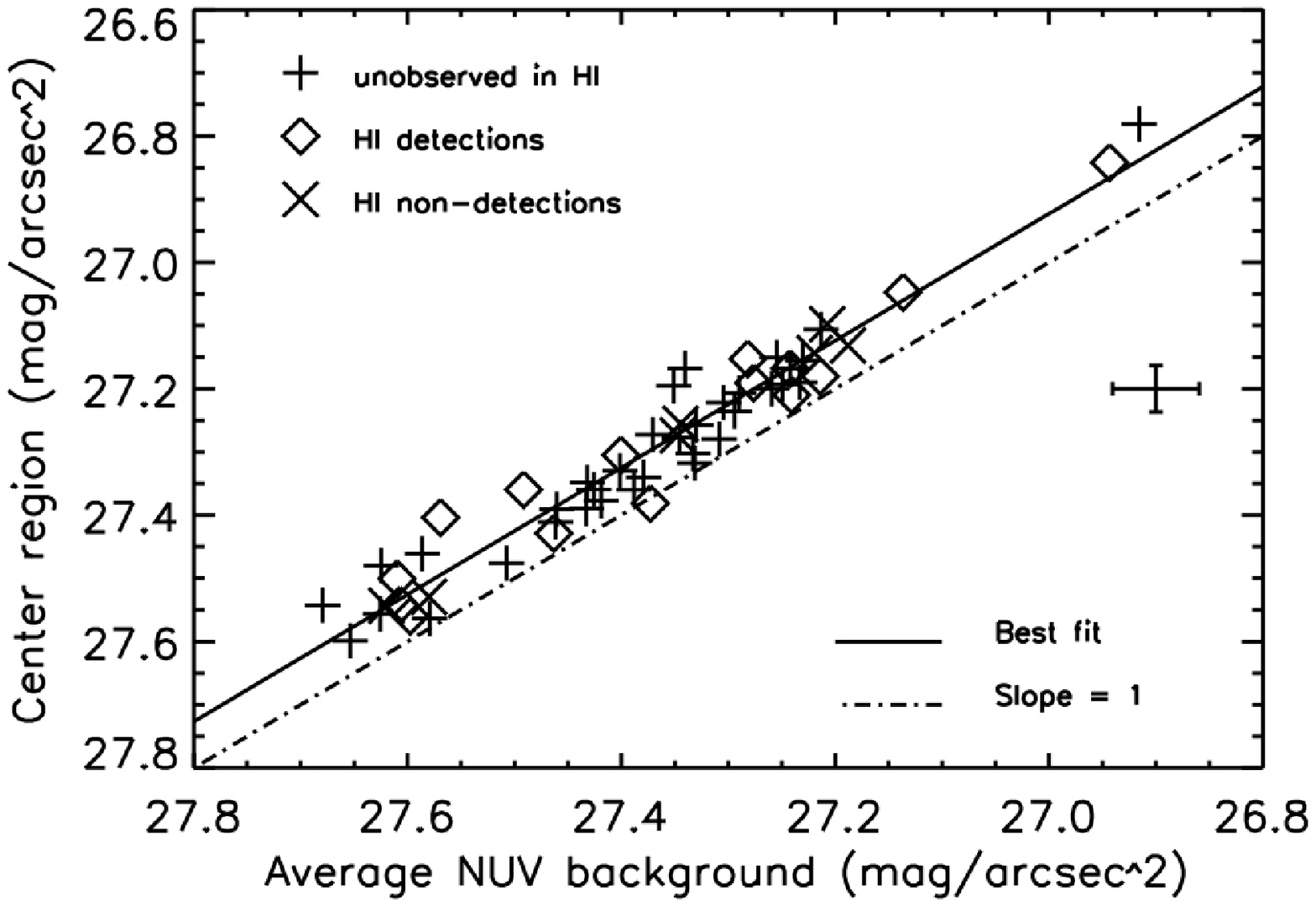}{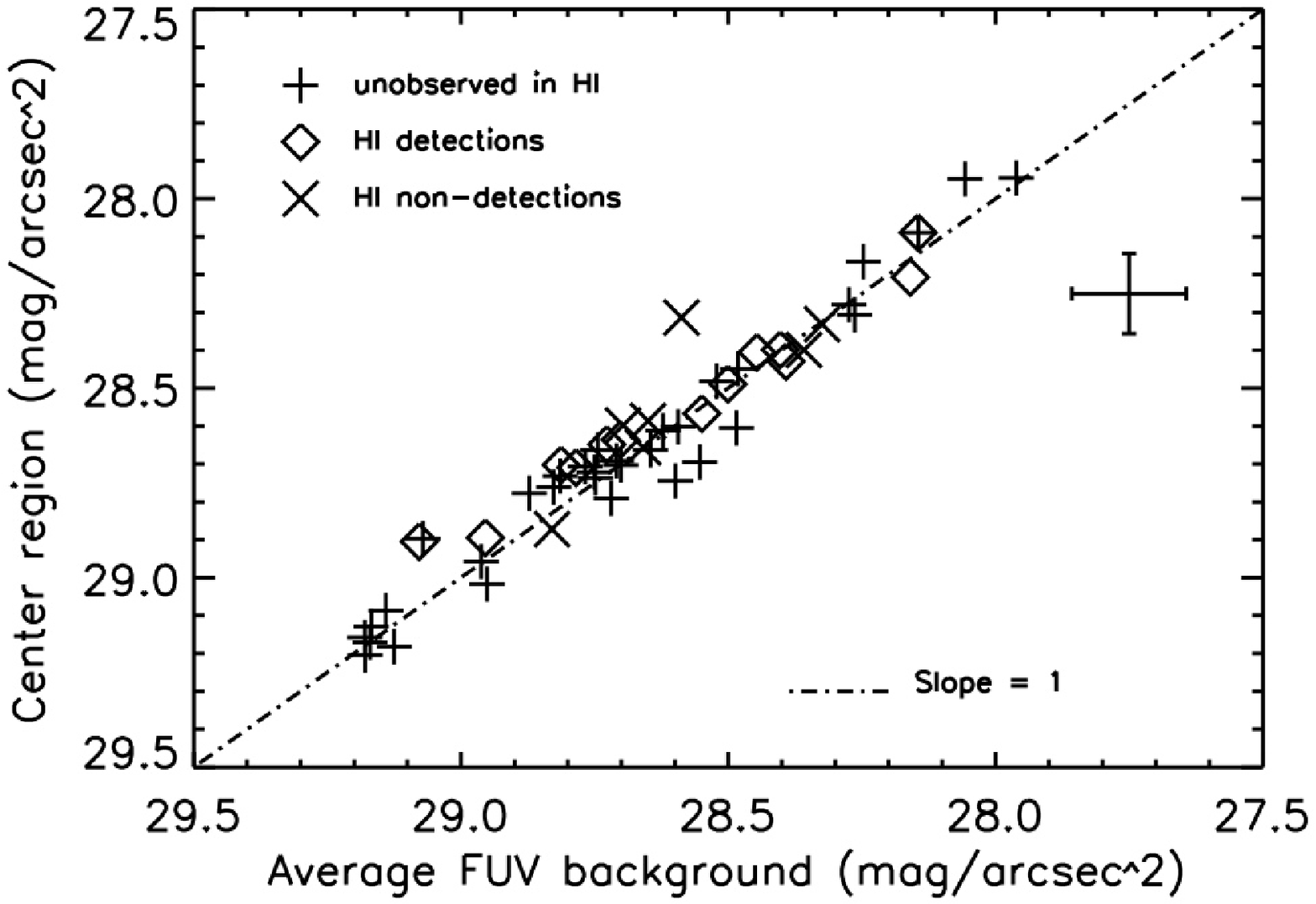}
\caption{Excess NUV emission (left) is seen in the environments of early-type galaxies; corresponding FUV excesses (right) are not detected. \label{fig:excess} Galaxies which have been observed in HI are plotted with different symbols; HI non-detections are at the level of a few $\times$ 10$^{7}$ M$_{\odot}$. Typical error bars are shown. No correlation is apparent with HI detections or non-detections.}
\end{figure}

\noindent $\it{NUV~Excess~vs.~M_{B}:}$ $\Sigma_{SFR}$ implied by ultraviolet excesses, or average NUV counts in the region centered on the galaxy minus average NUV counts in the background, are shown in Figure~2 (left) as a function of the absolute B magnitude of each galaxy. Error bars are derived from the scatter in the background regions. The SFR is calculated via SFR (M$_{\odot}$ yr$^{-1}$) = 1.4 $\times$ 10$^{-28}$ L$_{v}$ (ergs s$^{-1}$ Hz$^{-1}$) \citep{Kennicutt98}. The dot-dashed line indicates $\Sigma_{SFR}$ = 0. Brighter galaxies exhibit a higher ÒfaintÓ NUV excess and thus a higher $\Sigma_{SFR}$. The various symbols indicate HI detections and non-detections (as above).

\begin{figure}
\plottwo{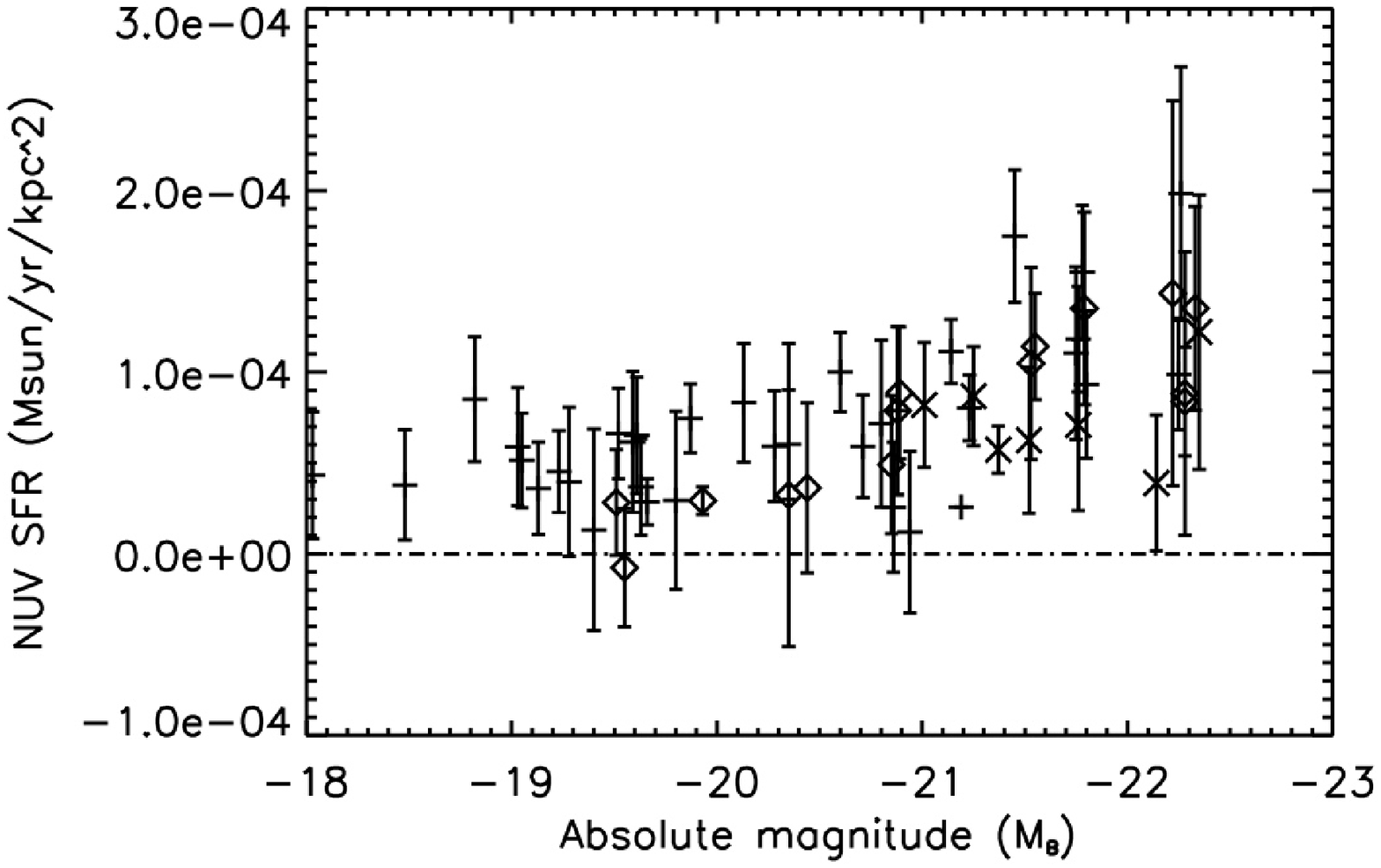}{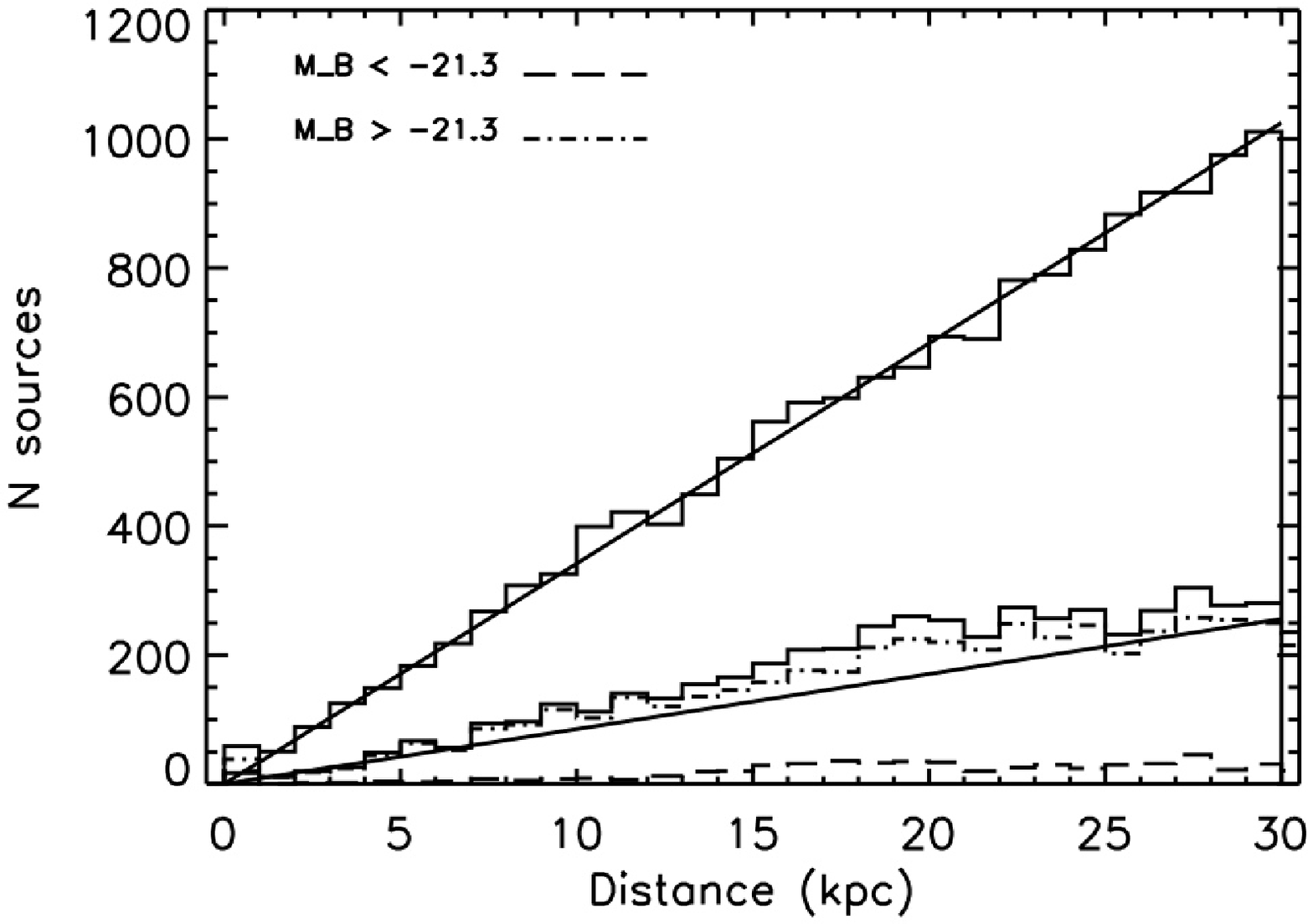}
\caption{At left, brighter early-type galaxies exhibit higher NUV excesses, interpreted here as higher star formation rates. At right, the distribution of $\geq$5$\sigma$ sources is shown. \label{fig:sfr} The steeper solid histogram and corresponding linear fit indicate the distribution of distances from the center of the background ``sky" regions. The shallower solid histogram (and the background ``sky" fit adjusted for the smaller area of the central region) indicate the distribution of distances from the galaxy in the center region, which is further broken down into ``bright" and ``faint" early types. }
\end{figure}

\noindent $\it{Bright~UV~Excesses:}$ All sources detected at or above 5$\sigma$ within 30 kpc of each galaxy, as well as within the regions defined as background ``sky", are shown in the histograms in Figure~2 (right). 
The ``sky" regions comprise four times as much area on each GALEX tile as the central region, but there are disproportionately more sources in the central regions. The fit to the ``sky" region distribution shows a linearly increasing trend with distance, but there is an excess of sources in the central region compared to a corresponding linear fit to its decreased area. Again, this excess is inconsistent with being due to background sources. Further, we separate the sources within 30 kpc of bright early types (defined as having MB $<$ -21.3, 19 systems) from those within 30 kpc of faint early types (defined as having MB $>$ -21.3, 37 systems). Faint early types, comprising 2/3 of the sample, are surrounded by four times as many sources of at least 5$\sigma$ compared to bright early types, which comprise the other 1/3 of the sample. The peak of the distribution of sources surrounding the bright early types also tends to be brighter and slightly redder than those found around faint early types. The spatial distribution of sources around all early type galaxies increases approximately linearly out to 20 kpc, where the distribution flattens out to 30 kpc (Figure 2).

\section{Conclusion}

The excesses described in this paper are not examples of the UV upturn phenomenon (believed to be caused by old, hot, helium-burning stars), where the FUV is by definition brighter than the NUV, as our excesses are not detected in the FUV. We believe that we are detecting the extended envelopes of early-type galaxies in a more extended, mostly diffuse star forming component than has been detected previously, though the excess of bright NUV sources indicates that stars are forming in clumps as well. Though these clumps are possibly correlated with known stellar clusters, preliminary results indicate that we generally do not detect bright UV sources where globular clusters have been catalogued in our sample galaxies.

We detect NUV excesses in the outskirts of our entire early-type galaxy sample. Faint NUV excesses increase with galaxy luminosity and are not correlated with the presence or absence of HI in the environments of these galaxies. The faint NUV excesses in the outskirts of early-type galaxies can be interpreted as being due to the presence of star formation at or above a few $\times$ 10$^{-5}$ M$_{\odot}$ yr$^{-1}$ kpc$^{-2}$; star formation at this rate can create a few percent of the mass of an early-type galaxy in a Gyr. Faint early types, comprising two-thirds of the sample, exhibit four times as many bright UV sources within 30 kpc compared to bright early types (which constitute the other third of the sample), and the sources detected around faint early types tend to be fainter and bluer than those detected around bright early types. This indicates that early types with M$_{B}$ $>$ -21.3 are more likely to be building up their mass with young stars. The spatial distribution of bright sources around all early types increases approximately linearly out to 20 kpc and subsequently flattens out to 30 kpc. 

\bibliography{up2010_ref}

\end{document}